\documentstyle[pra,aps]{revtex}
\renewcommand{\bf}[1]{\mbox{\boldmath{$#1$}}}
\begin{document}
\draft

\title {Positronium in crossed electric and magnetic
fields: the existence of a long-lived ground state}

\author{J. Shertzer}
\address{College of the Holy Cross, Worcester MA 01610}

\author{J. Ackermann}
\address{Institut f\"ur molekulare Biotechnologie (IMB),\\
 Beutenbergstrasse 11,
D-07745 Jena, Germany}

\author{P. Schmelcher}
\address{Theoretische Chemie, Physikalisch-Chemisches Institut, Universit\"at
Heidelberg, \\
Im Neuenheimer Feld 253, D-69120 Heidelberg, Germany}

\date{\today }
\maketitle

\begin{abstract}
It was earlier reported [PRL {\bf 78} 199, (1997)]
that long-lived excited states of positronium
can be formed  in  crossed electric and 
magnetic fields at laboratory field strengths.
Unlike the lower-lying states that
are localized in the magnetically distorted Coulomb well,
these long-lived states which can possess a lifetime
up to many years are localized in an outer potential well
that is formed for certain values of the pseudomomentum
and magnetic field.
The present work extends the original analysis and studies
the dependence  of the spectrum  as a function
of  field strength and
pseudomomentum over a wide range of parameters.  
We
predict that  in the limit of large pseudomomentum,
the ground state of positronium atom in a magnetic field
will become delocalized; for strong fields, the
binding energy of this state is quite large,  resulting in
a  ground state that is both stable 
against direct annihilation and against ionization
by low frequency background radiation.  

\end{abstract}

\section {Introduction}

The problem of treating the two-body system in a magnetic
field has a long history
\cite {Rau,Bhat,Clark,Burkova,Gay79,Fauth,Baye,Vincke,Farrelly}.  
Early on it was realized that the
center of mass motion cannot be separated from the internal motion.
The motional electric field due to the collective
motion of the system in the magnetic field gives rise to
a Stark term in the Hamiltonian which is very similar to
that arising from  an external electric field.
Hence, both problems can be treated in a unified way.

When the center of mass effects are treated correctly,
the total momentum of the system ${\bf P}$ is not a  conserved quantity and a separation of the center of mass and internal motion is impossible.   The
pseudomomentum ${\bf K} = {\bf P} + {e\over 2} 
{\bf B} \times {\bf r}$ is however a conserved  quantity,
where ${\bf P}$ is the total momentum and
${\bf r}$ is the relative vector of the two
oppositely charged particles with charges ${\pm e}$\cite{Av78}.
For  neutral systems, one can carry out a pseudoseparation
of the center of mass and internal motion, where the
effective Hamiltonian for the internal motion includes 
$K$-dependent terms, where $K$ is the eigenvalue of the pseudomomentum
\cite{Av78,Jo83}.  
Recently, it was shown that one can account for the effect
of the motion of the center of mass on the internal motion via an effective
potential that is gauge independent\cite{Di94}.  This potential
gives rise to an outer well for certain values of
$K$ and $B$, which leads to
delocalized states. There have been several studies of the
effects of a significant Stark term on the strong field
behavior of one-electron atoms  \cite{BP94,P94,B95}.

Recently, the gauge-invariant potential was extended to the case of
positronium, where rigorous numerical results indicated
that for  field strengths attainable in the laboratory,
one could form delocalized states of positronium\cite{us}.
The probability density within an interparticle distance radius of
hundreds  of Angstroms is less than $10^{-16}$, indicating that
the rates for direct annihilation and radiative decay to lower lying
states localized in the Coulomb well are near zero. 
The existence of such delocalized
states for positronium is important because it
provides a mechanism for creating a stable bound state
of a particle-antiparticle pair.  

Our previous work  established the existence of
such a state at values of $K$ and $B$ that are
attainable in the laboratory.  In this paper, we 
attempt a more thorough investigation of the problem.
In section II, we describe the potential surface
of positronium (Ps) as a function of  $K$
for values of the magnetic field from laboratory
field strengths ($10^{-5}-10^{-4}$a.u.)  up to superstrong fields
(10-100 a.u.).  We discuss qualitatively the effect
of this potential on the energy spectrum.
We present 
numerical results for select values of the parameters
$K$ and $B$ and 
show that for large values of the pseudomomentum 
the ground state of Ps becomes a delocalized state.
This is an important result because it means that 
any Ps formed under the proper conditions will  be
stable against annihilation.  
In the conclusion, we speculate on how one may obtain
evidence of long-lived Ps in a laboratory.

\section{The potential surface and energy spectrum
of  positronium as a function of
B and K}  

Throughout this paper we use atomic units:
\begin{quasitable}
\begin{tabular}{lll}

Charge   &$e$                           &$4.8029\times 10^{-10}$e.s.u.\\
Mass     &$m$                           &$9.1085\times 10^{-28}$gm \\
Length   &$a_o=\hbar^2/me^2$            &$5.2917\times 10^{-9}$cm\\
Velocity &$v_o=\alpha c=e^2/\hbar$      &$2.1877\times 10^{ 8}$cm/s\\
Momentum &$p_o=m\alpha c=me^2/\hbar^2$  &$1.9926\times 10^{-19}$gm cm/s \\
Energy   &$me^4/\hbar^2$                &$4.3590 \times 10^{-11}$ erg\\
Magnetic field
         &$m^2e^3/\hbar^3$                &$2.350 \times 10^9$G\\
Electric field
         &$m^2e^5/\hbar^4$              &$5.142 \times 10^9$ V/cm\\
\end{tabular}
\end{quasitable}
The effective Hamiltonian for the positronium atom in crossed
electric and magnetic fields which results from the gauge invariant
pseudoseparation \cite{Di94} is
\begin{equation}\label{eq:H}
H={{\bf p}^2\over 2}+{1\over 4} ({\bf K} + {\bf B} \times {\bf r})^2
- {1\over r} + {\bf E} \cdot {\bf r}
\end{equation}
where $T={{\bf p}^2\over 2}$ is the kinetic energy and
$V={1\over 4} ({\bf K} + {\bf B} \times {\bf r})^2 
-{1\over r} + {\bf E}\cdot {\bf r}$ is the potential of the
relative motion.  ${\bf K}$ is the eigenvalue of the conserved
pseudomomentum.

The term in the potential that depends explicitly on the
external electric field can be eliminated by defining 
an effective pseudomomentum 
${\bf K'}={\bf K}-2{\bf v_d}$, where ${\bf v_d} = {\bf E} \times {\bf B}/B^2$ is the classical drift velocity of a free charged particle in crossed fields. Using this definition in
Eq. (\ref{eq:H}) ,  we have
\begin{equation}\label{eq:HK'}
H={{\bf p}^2\over 2} +{1\over 4}({\bf K'} + {\bf B} \times {\bf r})^2
-{1\over r} + {\bf K'}\cdot {\bf v_d} + {\bf v_d}^2 
\end{equation}
It is now clear that the  energy eigenvalues  of positronium
in an external electric field can be obtained from
the zero external electric field results by replacing ${\bf K}$ with 
${\bf K'}$
and shifting the energy ${\cal E}$ by a constant amount:
\begin{equation}\label{eq:energy}
{\cal E}({\bf K},{\bf B},{\bf E}) = {\cal E}({\bf K'},{\bf B}, 0)
+{\bf K'}\cdot {\bf v_d} + {\bf v_d}^2 
\end{equation}

It is important to note that the ionization threshold $I$,  which
is the zero point energy of the free particles in the presence
of the fields, is different in the two cases.  In a pure
magnetic field $I=B$. In crossed fields, the ionization threshold
includes the energy due to the drift motion  $I=B+
{\bf K'} \cdot {\bf v_d} +{\bf v_d^2}$.
The ionization threshold is shifted by the same
amount as the energy levels.
The external electric field can be used to control the
value of the effective pseudomomentum ${\bf K'}$ experimentally.
For 
any combination of ${\bf E}$ and ${\bf K}$ that
leaves the effective pseudomomentum ${\bf K'}$
unchanged, the ionization energy
$I-{\cal E}$ is a constant.

Without loss of generality, 
we choose ${\bf B} = B\hat z$, 
${\bf E}=0$, and ${\bf K}=K\hat y$.  
Components of ${\bf K}$ parallel to ${\bf B}$ shift the energy
by a constant amount.
The effective Hamiltonian for the internal motion 
of positronium
is given in Cartesian coordinates by
\begin{equation}\label{eq:Hcart}
H={\bf p}^2 +
{B^2\over 4}(x^2+y^2) + {BKx\over 2}
-{1\over \sqrt{x^2+y^2+z^2} } +  {K^2\over 4}\quad.
\end{equation}
The Hamiltonian for other particle-antiparticle pairs can be
obtained through a simple scaling of the Ps Hamiltonian.
From the eigenvalues of the Hamiltonian   for Ps ${\cal E}(K,B)$,
we can obtain results for the particle-antiparticle pair
with   $m_1=m_2=m$ (m is a dimensionless scale parameter) as follows:
\begin{equation}\label{eq:massscale}
\tilde{\cal E}(m,\tilde K,\tilde B)= m{\cal E}(K={\tilde K\over m},
B={\tilde B\over m^2}) .
\end{equation}

All terms in the potential increase with increasing
$\vert x\vert$, $\vert y \vert$,  and $\vert z \vert $ with the exception of
${BKx\over2}$ which decreases for negative $x$.
In addition to the minimum at the Coulomb singularity
$V(0,0,0)=-\infty$, for sufficiently large $K$ the potential has another local minimum $V_o=V(x_o,0,0)$
where $x_o<0$ 
(see Fig.~1).
To determine
the critical value $K_c$ for which the outer well on the negative
$x-$axis appears,
we set $\partial V(x,0,0)/\partial x$=0:
\begin{equation}\label{eq:cubic}
x^3+{K\over B} x^2 - {2\over B^2} = 0 , \quad x<0.
\end{equation} 
If $K\ge K_c=^3\!\sqrt{{27B\over 2}}$, the roots of the equation are real and given by: 
\begin{eqnarray}\label{eq:roots}
x_1&=&{K\over 3B}\biggl[2\cos\biggr({\theta\over 3}\biggr)-1\biggr]\nonumber\\
x_2&=&{K\over 3B}\biggl[2\cos\biggr({\theta+2\pi\over 3}\biggr)-1\biggr]\\
x_3&=&{K\over 3B}\biggl[2\cos\biggr({\theta+4\pi\over 3}\biggr)-1\biggr]\nonumber
\end{eqnarray}
where $\cos\theta={27B\over K^3}-1$. The angle is bound
by    $0\le\theta\le \pi$.
$\theta=0$   corresponds to the critical value $K=K_c$; for
$K> K_c$, the potential surface has two local minima.
$\theta=\pi$ corresponds to  ${B\over K^3} \rightarrow 0$
or $K\rightarrow \infty$. 
It is clear from the
form of the solutions that
$x_2 \le x_3 < 0 < x_1$; $x_1$ is unphysical because Eq.~(\ref{eq:roots})  holds
only for $x<0$.  
The location of the saddle $x_s$ and outer well
minimum $x_o$ are given by
\begin{eqnarray}\label{eq:xoxs}
x_s&=&x_3=-{K\over3B}\biggl[1+\cos\biggl({\theta\over3}\biggr)-\sqrt{3}
\sin\biggl ({\theta\over 3}\biggr )\biggr ] \nonumber\\
x_o&=&x_2=-{K\over3B}\biggl[1+\cos\biggl({\theta\over3}\biggr)+\sqrt{3}
\sin\biggl ({\theta\over 3}\biggr ) \biggr ]
\end{eqnarray} 

We now discuss the potential surface, the ground state,
and energy spectrum  for a fixed
$B$ as a function of $K$. We will hereafter refer to
the potential well that includes the origin as the
magnetically distorted Coulomb well (MDCW).  
The additional  well that forms on the negative $x-$axis
will be referred to as the outer well (OW).    
The ionization threshold for positronium
is given
by the field strength $I=B$.  
All numerical results were obtained by applying the adaptive
3D finite element method (FEM) \cite{Ack93} to the solution of the Schr\"odinger
equation for the Hamiltonian given in Eq.~(\ref{eq:H}).

\subsection{$K=0$}
At $K=0$, the Hamiltonian simplifies to
\begin{equation}\label{eq:K=0}
H={\bf p}^2 +
{B^2\over 4}(x^2+y^2)  
-{1\over \sqrt{x^2+y^2+z^2} } 
\end{equation}
where ${\bf p}^2-{1\over r}$ is the Hamiltonian for Ps in the field free case.   The field dependent term ${B^2(x^2+y^2)\over4}$ destroys the
rotational symmetry of the field-free Hamiltonian and
the  total angular momentum $L$ is no longer
a good quantum  number.
The azimuthal
symmetry of the problem remains and the z-component of
the angular momentum is still a good quantum number.
The field dependent term  provides additional confinement
in the $x$ and $y$ directions.

The  field dependent term in the  Hamiltonian is positive definite
and thus the energy eigenvalues of the field free Hamiltonian
are increased by the non-zero $B$ field.
For the ground state, this effect is offset by
the increase in the ionization threshold:  the  
binding energy of the ground state increases
with increasing $B$ field (see Table 1).  For high lying states,
the effect of the additional confinement  perpendicular
to the field is more pronounced
and  Rydberg states of the field free problem may be pushed
into the continuum by the field.  

The energy levels in the MDCW can be obtained  by applying well-known
scaling relations to the hydrogenic results in the infinite
mass approximation which are available in the literature
\cite{Rosner,Ruder,KLJ}.
From the results  $\tilde{\cal E}(\tilde B)$ for hydrogen (with $S=0$ and $L_z=0$),  
we can obtain results for Ps with zero pseudomomentum:
\begin{equation}\label{eq:hy}
{\cal E}(K=0,B)= {1\over 2}\tilde{\cal  E}(\tilde B=4B)
\end{equation}

\subsection{$0\le K\le K_c$} 
As $K$ is increased  monotonically from $0$ 
up to  the critical value, we analyze the effect
of the two $K$-dependent terms  in the potential  for a fixed $B$.  
The term $K^2/4$ shifts the 
entire potential curve upward by a constant amount.
The  asymmetric
term
${BKx\over 2}$
destroys the azimuthal symmetry for any $K>0$ and further distorts
the shape of the well.  

The combined effect of both terms 
leads to a decrease in
the binding
energy of the ground state for fixed $B$ 
with increasing $K$ (see Table 1).  
Second order perturbation theory can be used
to calculate the energy shift of the ground state
when both $B$ and $K$
are small.
The average interparticle distance $\langle x \rangle$ 
for the ground state decreases
($\vert \langle  x \rangle \vert$ increases) 
with increasing $K$, indicating a `decentering' of
the probability density.  This effect can be quite
dramatic for large $B$, where $K<K_c$ does not
imply small $K$ (see Table I).  

For high lying states, 
the asymmetry of the well becomes even more important
resulting in 
greater decentering of the probability 
density than what is  observed for the ground state.
Because of the competing effects of the two $K$-dependent
terms,
it is impossible
to predict the net effect on the energy levels
near the ionization threshold. A full 3D solution of
the Schr\"odinger equation is required for each set of
parameters.

\subsection{$K=K_c$}

The critical value $K_c$ is the largest value of $K$ for
which only a single minimum exists in the potential surface.
At the critical value  $K_c$
the saddle point $x_s$ and outer well minimum  $x_o$ coincide at
\begin{equation}\label{eq:xc}
x_c=-{9\over  K_c^2}=-\biggl ({2\over B}\biggr) ^{2\over3} 
\end{equation}
The potential at 
this point is 
\begin{equation}\label{eq:Vc}
V_c=V(x_c,0,0)=-{K_c^2\over 12} = -{3\over 12}\biggl ({B\over 2}\biggr )^{2\over 3}
\end{equation}
For laboratory field strengths ($B\approx10^{-5}$), the
magnitude of the critical point is on the order of several thousand Angstroms;
for strong fields ($B\approx 1$), the critical point is
one the order of 1 Angstrom.

\subsection {$K_c  < K$}

\subsubsection{General features of the potential}

If we now allow $K$ to further increase 
above the critical value, we observe the formation of
the OW, which is separated from the MDCW by the saddle.
The outer well minimum is denoted $V_o=V(x_o,0,0)$
and the height of the saddle is $V_s=V(x_s,0,0)$.
For fixed $B$, we observe 
the following general features of the potential:

1. The value of 
$x_o$  decreases 
with increasing K; the outer well moves away from
MDCW along the negative x-axis 
($ x_c\ge x_o\ge -\infty$).

2. The minimum of the outer well
$V_o$ increases with increasing $K$ but is bounded from above by zero
($V_c\le V_o \le 0$ ). 

3. The value of 
$x_s$ increases with increasing  $K$; the barrier maximum  moves
towards the origin 
($x_c\le x_s \le 0$). 

4. The height of
the  saddle $V_s$
increases without limit with increasing $K$ ($V_c\le V_s\le \infty $).

In Fig.~2, we show  $|x_o|$ and $|x_s|$ as a function
of $K$ for $B=10^{-4}, 10^{-3},
10^{-2}, 10^{-1}, 1$ and $10$ on a 
log-log plot; the point of coalescence is the
critical value $x_c$. Note that the curves are identical
in shape.
In Fig.~3, we show $V_o$ and $V_s$ as a function of $K$ at the same field
strengths as Fig.~2; the point of coalesence is the value
of the potential at the critical point, $V_c=V(x_c,0,0)$.
The shapes of the curves are  similar, with $V_s$ rising
dramatically for $K>K_c$;  $V_o$ approaches zero
in the limit $K\rightarrow \infty$. 
 
It is also important to note that the shape of the MDCW
changes above the critical value of $K$.  For $K>K_c$
and $\vert x\vert <0.75\vert x_s\vert $,
${BK|x|\over 2}<{K^2\over 4}$ and the potential
V(x,0,0) near the origin rises sharply, resulting
in a narrow well 
(see Fig. 1). 
The outer well , in contrast,
becomes broader and deeper with increasing $K$.
Although the well minimum is increasing (approaching zero
from below) the barrier height is increasing at a much
faster rate than the minimum is increasing.

Despite the similarity of the potential 
for different
values of $B$ (see Fig.~1,3) the
energy spectrum of the system - including the ground
state - is
dramatically different depending on the value of $B$.
The bound states of the system depend not just on
the potential, but also on the ionization threshold $I$.

When the two wells are distinct, that is, the  saddle point energy
is well above the ionization threshold, the bound states
can be classified as OW states or MDCW states: their
probability density is concentrated solely in one well.  
The bound states of the OW are called delocalized states
because the average interparticle separation is large.
The
overlap between an OW state and a MDCW state is essentially zero.
All of the  states that are localized in the MDCW are
pushed upward in energy with increasing $K$  and into the continuum;
the number
of MDCW bound states decreases. 
The number of bound states in the outer well increases
with increasing $K$.

However, before we reach the regime where the problem reduces
to two separate wells, there is a range of $K$ for which
{\it the saddle point energy is less than the ionization threshold.}
If the energy of a state is considerably less than
$V_s$, the state will  be concentrated in the OW
or the MDCW with very little probability in the barrier region
and the above classification scheme for states is still valid.
However, if the energy of a state is close to  $V_s$, 
the distinction between an OW state and an MDCW states becomes murky.
(It is important to remember that $V_s$ is the potential
maximum in the $x-$direction only; the ionization  threshold
$I=B$ as well as the energy of the state are quantities
that depend on the 3D wavefunction.)
Under these very special conditions, ${\cal E}\approx V_s < B$,
we have a third kind of state: the saddle state.

\subsubsection{Saddle states}

In the regime $K > K_c$, but $V_s < I$,  an interesting
phenomena can occur:  a MDCW state can undergo a continuous
transformation into an OW state with increasing $K$
as  the energy of the MDCW state approaches $V_s$.
The MDCW state (see Fig. 4a) will begin to leak into the
classically forbidden barrier region
(see Fig. 4b).  
The amount of tunneling through the barrier
increases with increasing $K$, until a saddle
state forms, which has  probability density in  the
MDCW, the barrier region and the OW (see Fig. 4c).
At even higher $K$, more and more probability density
is transfered from the MDCW to the OW (see Fig. 4d). Since the barrier
height is increasing,  the energy of the
state is eventually less than $V_s$ and  
the state is fully delocalized in the OW (see Fig. 4e). 
The range of $K$ over  which
this process occurs can vary orders of magnitude
depending on the relative values of $I$, $V_s$ and ${\cal E}$
as a function of $K$ (see Table 1).

For low magnetic fields, the gap between the ionization
threshold and the saddle point energy is already small at
the critical value; hence, the formation of saddle
states is limited to a very small range of $K$.
For $B=10^{-4}$,  the critical value is  $K_c=
0.1105$ and saddle point reaches the ionization threshold
($V_s=I$)
at $K=0.1414$.  Saddle states can only form
over a range $\Delta K \approx .03$. Many bound states still exist in the
MDCW when  the  saddle point reaches the ionization threshold.
In contrast, for high magnetic fields, the 
separation between the ionization threshold and the
saddle point energy is orders of magnitude greater
than what it is at low fields.
For $B=1$,
the critical value is $K_c=2.381$ and
the saddle point reaches the ionization threshold at
$K=3.800$. Saddle states can form over a range
$\Delta K \approx 1.4$.  The ground state energy of the system
is itself a saddle state in this range.

Once the saddle point reaches the ionization threshold
for a fixed $B$,  saddle states can not be formed and 
all  of the MDCW states are pushed into the continuum
with increasing $K$;  the OW spectrum is
continuously changing with increasing $K$.

\subsubsection{Fate of the ground state}

It is clear that this process of pushing MDCW states out
of the well and into the continuum or  into the
OW  continues
until the MDCW spectrum is depleted of bound states.
The fate of the MDCW ground state is quite different for
low $B$ fields than for high $B$ fields.  At low fields,
there is a sudden  transition at which the ground
state  energy of the OW drops below the ground
state of the MDCW.  Both states are well below the
saddle point energy, which is much greater than
the ionization threshold.
(see Table 1, $B$=0.001, $K$=1.0, 1.1)
This is indicative of a sharp level crossing between
two distinct states.
The wavefunctions for these two states have zero overlap.
The MDCW ground state will continue to be pushed upward
in energy with increasing $K$ until it reaches the continuum.

In contrast, at high fields there is a slow migration
of the ground state wavefunction away from  the origin
and towards the outer well.  The  ground state energy
is greater than $V_s$ (but less than $I$) over a very wide range of $K$,
allowing for a slow transfer of probability with increasing $K$
as the MDCW state forms a saddle state which eventually
delocalizes.  
In  Fig. 5  we show the value of the ground state wavefunction
at the origin as a function of $K$  for several values of $B$.
Note for small $B$ the abrupt drop at the point of the level
crossing; for large $B$, there is a smooth  decay
indicative of the slow transfer of probability from
the MDCW to the OW.

The ground state of positronium will become
a delocalized state at sufficiently high $K$.
This is an important result, because it implies that
any positronium formed under these conditions will
have a long-lived ground state that is stable against
direct annihilation.  For low fields, the ground state is
extended in space and weakly bound;   at high field
strengths, 
the size of the atom in the high $K$ limit is small
and the binding energy is large.

The decay rate for positronium is
\begin{equation}\label{eq:gamma}
\Gamma = \sigma v \rho
\end{equation}
where $\sigma$ is the plane-wave cross section for
free pair annihilation, $v$ is the relative velocity of
the electron and positron, and $\rho$ is the square of
the wavefunction evaluated at contact. In the field-free case,
the lowest order decay rate for the ground state of 
parapositronium is $\Gamma =8.03 \times 10^{10} $s$^{-1}$;
for orthopositronium,  $\Gamma=7.21 \times 10^{6} $s$^{-1}$\cite
{Ad83,St75}.
The distinction between ortho- and para-positronium
is meaningless in the present  discussion.
The spin-spin interaction is dominated by
the spin-field interaction at field strengths $B>10^{-5}$. We assume the spins are aligned with the field and the
total energy and ionization energy are shifted by $B$.
Nevertheless,  the basic physics
contained in the decay rate formula is unchanged: the
decay rate depends on the probability density at the
origin and if the probability density 
is zero,  the Ps atom is stable against annihilation.

At low fields $10^{-5} < B < 10^{-2}$,
there is a well-defined crossing between the MDCW
ground state and OW ground state near $K=1$. The
probability density at the origin drops dramatically  at this
crossing -
over sixteen orders of magnitude - resulting in a lifetime on the order
of years. At higher fields, $0.1 < B < 10$, the probability density
at the origin decreases exponentially with increasing $K$ (Fig. 5);
for $B/K^2 < 10^{-2}$, the probability density is
less than $10^{-16}$, again resulting in a lifetime greater
than a year.

\subsection {$K_c << K$}

In order to extract more quantitative conclusions about the
spectrum of the outer well in the large $K$ limit, 
we expand Eq.~(\ref{eq:xoxs}) in powers of
$\epsilon$, where  $\epsilon=
 {B\over K^3} << 1$.  (Hereafter,  the high $K$ limit
implies $\epsilon <<1$.)
We obtain analytic  expressions
for the saddle and outer well quantities in the large $K$ limit,
retaining terms up to first order in $\epsilon$: 
\begin{eqnarray} \label{eq:highK}
x_o &\rightarrow & {-K\over B}\;\biggl[1-2\epsilon\biggr]=
{-K\over B}+{2\over K^2}\nonumber\\
V_o&\rightarrow&{-B\over K} \;\biggl[1+\epsilon\biggr] 
={-B\over K}- {B^2\over K^4}\nonumber\\ 
x_s &\rightarrow & {-K\over B} \;
\biggl [0+\sqrt{2\epsilon} + \epsilon \biggr ]=
-\sqrt{{2\over BK}}-{1\over K^2}\\
V_s&\rightarrow& {K^2\over 4} \;\biggl[1-\sqrt{2\epsilon}+2\epsilon\biggr]
={K^2\over 4} -\sqrt{32BK}+{B\over 2K}\nonumber
\end{eqnarray} 
These results have wider applicability than may be expected,
because $\epsilon$ is already small at the critical value:
$\epsilon(K=K_c)={2\over 27}$.  
Below we compare the exact values of the saddle and
OW parameters with the high $K$ limit approximate results
at $B=1$; the critical value is $K_c=2.381$.
\begin{quasitable} 
\begin{tabular}{llrrrr}
\tableline
$B$ &$K$   &\multicolumn{1}{c}{$x_o$ - High $K$} 
           &\multicolumn{1}{c}{$V_o$ - High $K$} 
           &\multicolumn{1}{c}{$x_s$ - High $K$}
           &\multicolumn{1}{c}{$V_s$ - High $K$}   \\ 

    &      &\multicolumn{1}{c}{$x_o$ - Exact}
           &\multicolumn{1}{c}{$V_o$ - Exact}
           &\multicolumn{1}{c}{$x_s$ - Exact}
           &\multicolumn{1}{c}{$V_s$ - Exact}   \\

1.0& 5 &-4.920  &-.20160  &-.672    &3.188  \\
   &   &-4.917  &-.20165  &-.680    &3.195\\
1.0 &10 &-9.9800  &-.100100 &-.4572  &20.578 \\
    &   &-9.9799  &-.100100 &-.4578  &20.579 \\
1.0 &20 &-19.99500 &-.05000625 &-0.31873  &93.7004\\
    &   &-19.99500 &-.05000625 &-0.31878  &93.7006\\
\tableline
\end{tabular}
\end{quasitable}

For large $K$,
the energy of the bound states in the outer well are
bounded from above and below by the well minimum and
the ionization threshold respectively:
$V_o< {\cal E}_n^{OW} < B$ ,
where $n$ refers to the set of   quantum numbers
used to label the OW states.  In order to make estimates
for the ground state, we
use the anisotropic harmonic oscillator (AHO)approximation
for the OW \cite{Di94}.  This approximation involves
an expansion of the Coulomb potential around the minimum,
retaining terms to order $x_o^{-3}$;  the AHO approximation
is valid only when $|x_o|$ is large and only for
states  such that $\langle z^2+y^2\rangle << x_o^2$.  
(For $B>1$, the high $K$ limit
does not
imply that the AHO approximation is valid: ${B\over K} <<1$ is a stronger
requirement than ${B\over K^3} << 1$.)

The AHO energy spectrum is given by
\begin{equation}\label{eq:AHO}
{\cal E}_{n_xn_yn_z} = (n_x+{1\over 2}) \omega_x
                    +(n_y+{1\over 2}) \omega_y+(n_z+{1\over 2})
                     \omega_z + C.
\end{equation}
where
\begin{equation}\label{eq:omega}
\omega_x = \sqrt {(B^2+{4\over x_0^3}) };~~~~ 
\omega_y = \sqrt {(B^2-{2\over x_0^3}) };~~~~
\omega_z = \sqrt {     {-2\over x_0^3} };~~~~
C = {2\over x_0}-{B^2 x_0^2\over 4} + {K^2\over 4}\nonumber
\end{equation}
In the high $K$-limit, 
$\omega_x \rightarrow B-2\epsilon$, $\omega_y \rightarrow B+\epsilon$,
and  $\omega_z
\rightarrow B\sqrt{2\epsilon}$ and the ground state energy is:
\begin{equation}\label{AHOhighK}
{\cal E}_g \rightarrow 
   B \biggl [ 1+ \sqrt{ {\epsilon\over 2}} 
-{\epsilon\over 2}\biggr ] -{B\over K} 
\biggl [ 1 + 5\epsilon \biggr ] 
\end{equation}
We  compare below the approximate formula
for the ground state
in the outer well
with the  results from a 3D finite element calculation.
\begin{quasitable}
\begin{tabular}{rrll}
\tableline
\multicolumn{1}{c}{$B$} &\multicolumn{1}{c}{$K$}
    &\multicolumn{1}{c}{${\cal E}_g$ - FEM }
    &\multicolumn{1}{c}{${\cal E}_g$ - AHO}\\
0.01   &2.0   &.00523    &.00521\\
0.01   &5.0   &.008061   &.008062\\
0.01   &10.0  &.0090222  &.0090222\\
\tableline
\end{tabular}
\end{quasitable}
We now turn to the spectrum  of OW excited states.
For low-lying states, the AHO is still valid.
Excitations in the z-direction dominate for large $K$.
The level spacing between the harmonic states 
($\omega_z=\sqrt{{2B^3\over K^3}}$) is decreasing at a 
faster rate than the energy gap  between the ionization threshold and the ground state
($I-{\cal E}_g={B\over K}$):
the density of
low-lying bound states in the OW increases 
with increasing $K$.
The AHO approximation will break down  long before we
reach the ionization threshold, although 
the potential remains  approximately separable and
harmonic in the $x$ and $y$.
In the $z$ direction, the level spacing decreases as
one approaches the ionization threshold (at constant $K$) due to the 
${1\over \sqrt{(x_o^2 + z^2)} }$  potential.  
We expect the number of OW  bound states
is infinite at high but finite $K$.
\section{Conclusion and Outlook}

We have investigated the properties of the positronium atom
in crossed electric and magnetic fields for a broad range
of magnetic field strengths and values of the pseudomomentum.
The occurence of magnetically distorted Coulomb well states,
saddle states and outer well states   gives rise to a spectrum
that is rich and  unique.
An intensive numerical study via the
finite element approach allowed us to investigate the different regimes
and study the transitions between different 
types of quantum states as a function of pseudomomentum.

We believe that there is rigorous theoretical and numerical
evidence  to support the   prediction
that long-lived states of positronium 
exist in crossed electric and magnetic fields.  The near
zero probability for particle overlap 
prevents direct
annihilation.  For large
pseudomomentum ($B/K^3 << 1$), the ground state itself is
a long-lived state.

Positronium in crossed fields is not an exotic system.
Particle-antiparticle pairs occur in many different
physical situations.
Naturally occuring  magnetic fields
range from a few mGauss 
to 10$^{12}$  in neutron stars.
The answer to the question 
whether  Ps has ever
been produced under the appropriate conditions for
stability is in our opinion
yes.  Detection of the system is a more
difficult problem.

One possible means for creating stable positronium at
laboratory field strengths is to prepare Landau states
of e$^+$ and e$^-$  with low relative velocity in the
$z$ direction. The initial positions and velocities of
the two particles in the external fields define the conserved
pseudomomentum.  Once an OW bound state is formed,
the large dipole moment $ex_o$ of the OW state should make
detection easy.  The difficulty  is to detect the  positronium atom
before it collides with the container walls.   It is not
practical to use field gradients
for trapping, as the external fields must be
constant over the dimensions of the atom.  
Another possibility is to scatter positrons off  hydrogen in
crossed fields.  A judicious choice of experimental
parameters may enhance the cross section for positronium
formation in an OW state.

The search for stable positronium need not be limited
to the laboratory.  The  spectra from
neutron stars may  provide evidence of transitions between
outer well states.  These suggestions are  not meant to
confine the  search for this system, but only to
initiate a discussion that should challenge  experimentalists
in many subfields.

In conclusion, we believe that the existence of stable
Ps will be relevant in many areas of physics. 
One can only speculate at this time on the possible
applications of storing energy via a stable particle-antiparticle
pair.

The Bundesministerium f\"ur Bildung
und Forschung (JA) and  the National Science Foundation
(JS)  are
gratefully acknowledged for support.

\begin{table} 
TABLE I. The  location of the saddle point $x_s$,
the value of the potential at the saddle $V_s=V(x_s,0,0)$,
the location of the outer well minimum $x_o$,   the value
of the potential at the  outer well minimum $V_o=V(x_o,0,0)$,
the ground state energy  ${\cal E}_g$,  
and the expectation value  $\langle x\rangle$ for the ground state as a function of $K$ for  several values of $B$.
All quantities
are in atomic units.

\begin{tabular}{lrrrrrrr} 
\multicolumn{1}{c}{$B$}      
&\multicolumn{1}{c}{$K$}      
&\multicolumn{1}{c}{$x_s$}      
&\multicolumn{1}{c}{$V_s$}      
&\multicolumn{1}{c}{$x_o$}
&\multicolumn{1}{c}{$V_o$}
&\multicolumn{1}{c}{${\cal E}_g$}
&\multicolumn{1}{c}{$\langle x \rangle$}\\
\tableline
0.001  &0.0  &         &          &       &          &-.2500  & .0000 \\
       &0.1  &         &          &       &          &-.2475  &-.00720\\
       &0.2  &         &          &       &          &-.2400 &-.00360\\
       &0.3  &-100.00  & .0000    &-273.2 &-.00348   &-.2275  &-.00540\\
       &0.4  &-78.92   & .0131    &-386.6 &-.00254   &-.2100  &-.00720\\
       &0.5  &-68.04   & .0320    &-419.7 &-.00202   &-.1875  &-.00900\\
       &0.6  &-60.91   & .0562    &-594.3 &-.00167   &-.1600  &-.0108\\
       &0.7  &-55.72   & .0858    &-695.9 &-.00143   &-.1275  &-.0126\\
       &0.8  &-51.70   & .1206    &-796.9 &-.00125   &-.0900  &-.0144\\
       &0.9  &-48.46   & .1606    &-897.5 &-.00111   &-.04475 &-.0162\\
       &1.0  &-45.78   & .2058    &-998.0 &-.00100   &-.0000  &-.0018\\
       &1.5  &-36.97   & .5081   &-1499 &-.00067   & .00034 &-1499\\
       &2.0  &-31.88   & .9370   &-1999   &-.00050  & .00051 &-2000\\
       &3.0  &-25.93   &2.173    &-3000   &-.00033  & .00067 &-3000\\
       &5.0  &-20.04   & 6.150  &-5000 &-.00020   & .00080 &-5000\\
       &7.0  &-16.92   & 12.13  &-7000 &-.00014   &00086   &-7000\\
       &10.0 &-14.15   & 24.86  &-10000 &-.00010   &.00090   &-10000\\
\tableline
0.01   &0.0  &         &         &        &          &  -.2498 &  .0000\\
       &0.1  &         &         &        &          &  -.2473 & -.0179\\
       &0.2  &         &         &        &          &  -.2398 & -.0357\\
       &0.3  &         &         &        &          &  -.2273 & -.0537\\
       &0.4  &         &         &        &          &  -.2099 & -.0716\\
       &0.5  &         &         &        &          &  -.1874 & -.0897\\
       &0.6  &-23.37   &-.0092   &-52.84  &-.0176    & -.1600  &-.1079\\
       &0.7  &-20.00   & .0125   &-65.31  &-.0148    & -.1275  &-.1262\\
       &0.8  &-17.95   & .0405   &-76.59  &-.0128    & -.0901  &-.1446\\
       &0.9  &-16.50   & .0745   &-87.38  &-.0113    & -.0477  &-.1633\\
       &1.0  &-15.37   & .1140   &-97.91  &-.0101    &  .00053 &-98.10\\
       &1.1  &-14.47   & .1590   &-108.3  &-.0092    &  .00139 &-108.4\\
       &1.2  &-13.72   & .2095   &-118.6  &-.0084    &  .00210 &-118.7\\
       &1.5  &-12.04   & .3928   &-149.2  &-.0067    &  .00367 &-149.1 \\
       &2.0  &-10.27   & .8026   &-199.4  &-.0050    &  .00523 &-199.5\\
       &3.0  &-8.280   & 2.007   &-299.7  &-.0033    &  .00679 &-299.8\\
       &5.0  &-6.365   & 5.935   &-499.9  &-.0020   &  .00806 &-499.9\\
       &10.0 &-4.482   & 24.55   &-1000   &-.0010   &  .00902 &-1000\\
\tableline
0.1    &0.0  &         &         &        &          &-.2323  & 0.000\\
       &0.2  &         &         &        &          &-.2234  &-0.2259\\
       &0.4  &         &         &        &          &-.1969  &-0.4787\\ 
       &0.6  &         &         &        &          &-.1532  &-0.8070\\
       &0.8  &         &         &        &          &-.0938  &-1.363\\
       &1.0  &         &         &        &          &-.0241  &-3.177 \\
       &1.2  & -5.583  &-.0762   &-10.00  &-0.0900   &.0219    &-9.780\\
       &1.5  & -4.329 & .0537   &-13.98  &-0.0689    &.0410    & -14.16\\
       &2.0  & -3.479 & .3949   &-19.47  &-0.0507    &.0559    &-19.57\\
       &3.0  & -2.707 & 1.493   &-29.77  &-0.0334    &.0702    &-29.81\\
       &5.0  & -2.042 & 5.260   &-49.92  &-0.0200    &.0818    &-49.93\\ 
\tableline
1.0    &0.0  &	       &        &        &           &.3596   & .000\\
       &0.2  &         &        &        &           &.3633   &-.1253\\
       &0.4  &         &        &        &           &.3745   &-.2533\\
       &0.6  &         &        &        &           &.3925   &-.3868\\
       &0.8  &         &        &        &           &.4168   &-.5286\\
       &1.0  &         &        &        &           &.4464   &-.6815\\
       &1.5  &         &        &        &           &.5349   &-1.127\\
       &2.0  &         &        &        &           &.6247   &-1.668\\
       &2.5  &-1.281   &-.4092  &-2.000  &-0.4375    &.6970   &-2.255\\
       &3.0  &-1.000   &0.000   &-2.732  &-0.3481    &.7481   &-2.831\\
       &4.0  &-.7892   &1.310  &-3.866  &-0.2542    &.8095   &-3.910\\
       &5.0  &-0.6804  &3.195   &-4.917  &-0.2017    &.8451   &-4.493\\
       &6.0  &-0.6091  &5.624   &-5.943  &-0.1675    &.8688   &-5.960\\
       &8.0  &-0.5170  &12.06   &-7.969  &-0.1252    &.8991   &-7.977\\
      &10.0  &-0.4578  &20.58   &-9.980  &-0.1001    &.9178   &-9.985  \\ 
      &20.0  &-0.3188  &93.70   &-19.99  &-0.0500    &.9567   &-19.996\\
\tableline
10.0   &0.0  &         &        &        &           &8.600  &.000\\
       &0.4  &         &        &        &           &8.603  &-.0364\\
       &1.0  &         &        &        &           &8.622  &-.0912\\
       &1.5  &         &        &        &           &8.649  &-.1372\\
       &2.0  &         &        &        &           &8.685  &-.1838\\
       &3.0  &         &        &        &           &8.779  &-.2791\\
       &4.0  &         &        &        &           &8.890  &-.3772\\
       &5.0  &         &        &        &           &9.003  &-.4778\\
       &6.0  &-.2337   &-.9246  &-.5284  &-1.764     &9.109  &-.5800\\
       &7.0  &-.2000   &1.250  &-.6531  &-1.476     &9.202  &-.6829\\
       &8.0  &-.1795   &4.054   &-.7659  &-1.277     &9.281  &-.7857\\
       &9.0  &-.1650   &7.445   &-.8738  &-1.127     &9.346  &-.8881\\
       &10.0 &-.1537   &11.40   &-.9791  &-1.010     &9.400  &-.9901\\
       &15.0 &-.1204   &39.28   &-1.491  &-0.6687    &9.572  &-1.495\\ 
       &20.0 &-.1027   &80.26   &-1.995  &-.5006     &9.663  &-1.997\\
       &50.0 &-.0637   &593.5   &-4.999  &-.2000     &9.9474 &-5.000  \\
\tableline
100.0  &0.0  &         &        &        &           &97.07  &.000\\
       &1.0  &         &        &        &           &97.08  &-.00983\\
       &2.0  &         &        &        &           &97.09  &-.01965\\
       &4.0  &         &        &        &           &97.14  &-.03934\\
       &6.0  &         &        &        &           &97.22  &-.05907\\
       &8.0  &         &        &        &           &97.33  &-.00789\\
       &10.0 &         &        &        &           &97.45  &-.00987\\
       &15.0 &         &        &        &           &97.78  &-.01487\\
\tableline \end{tabular}
\end{table}
\newpage

\begin{figure}
\caption{$ V(x,0,0)$ at (a) $B=0.001$ for $K=0, K_c=.2381$, and  $K=.4$
and (b) $B=1.0$ for $K=0$, $K_c=2.381$, and $K=4.0$.
All quantities are in atomic units.}
\end{figure}

\begin{figure}
\caption{The location of the saddle point $x_s$ and the outer
well minimum $x_o$  as a function of $K$  for various values
of $B$ (in atomic units). The lowest point on each curve is  the critical 
point $x_c$. All quantities are in atomic units.} 
\end{figure}

\begin{figure}
\caption{The value of the potential at the saddle point
$V_s=V(x_s,0,0)$ and the OW minimum $V_o=V(x_o,0,0)$
as a function of $K$ for various values of $B$ (in atomic units).  
The saddle point point energy increases without limit
from the critical value $V_c=V(x_c)$;  the outer
well minimum increases from the critical value and
approaches zero in the limit $K \rightarrow \infty$.}
\end{figure}
\begin{figure}
\caption{The evolution of a MDCW state  into an OW state
as a function of $K$ at $B=0.1$;
(a)  MDCW state with minimal tunneling through the barrier;
(b)  significant tunneling of the MDCW state through the barrier;
(c)  saddle state;
(d)  decreased probability density in the MDCW;
(e)  a fully delocalized OW state.}
\end{figure}

\begin{figure} \caption{The value of the wavefunction at the origin as
a function of $K$ for various values of $B$ (in atomic units).}
\end{figure}

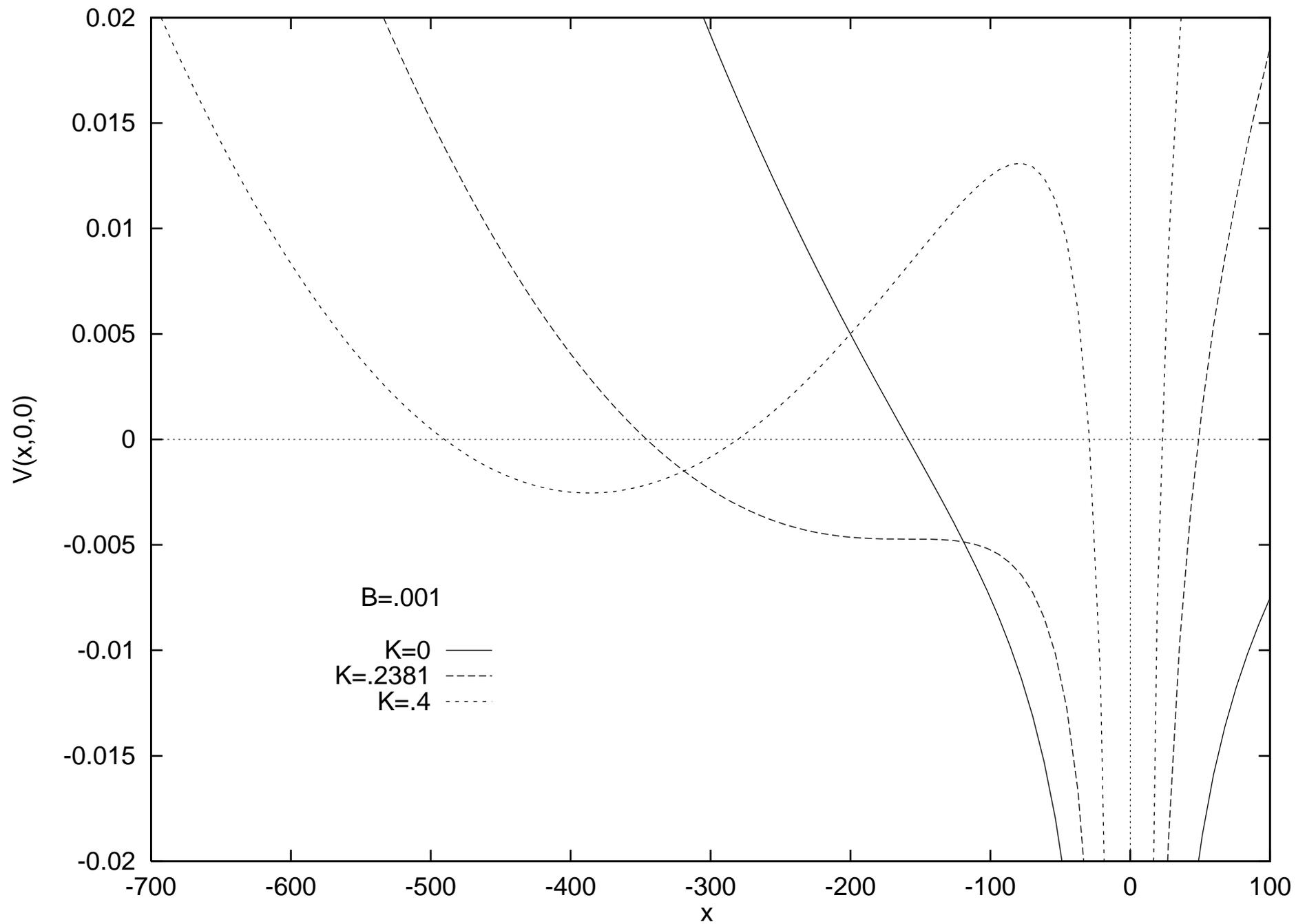
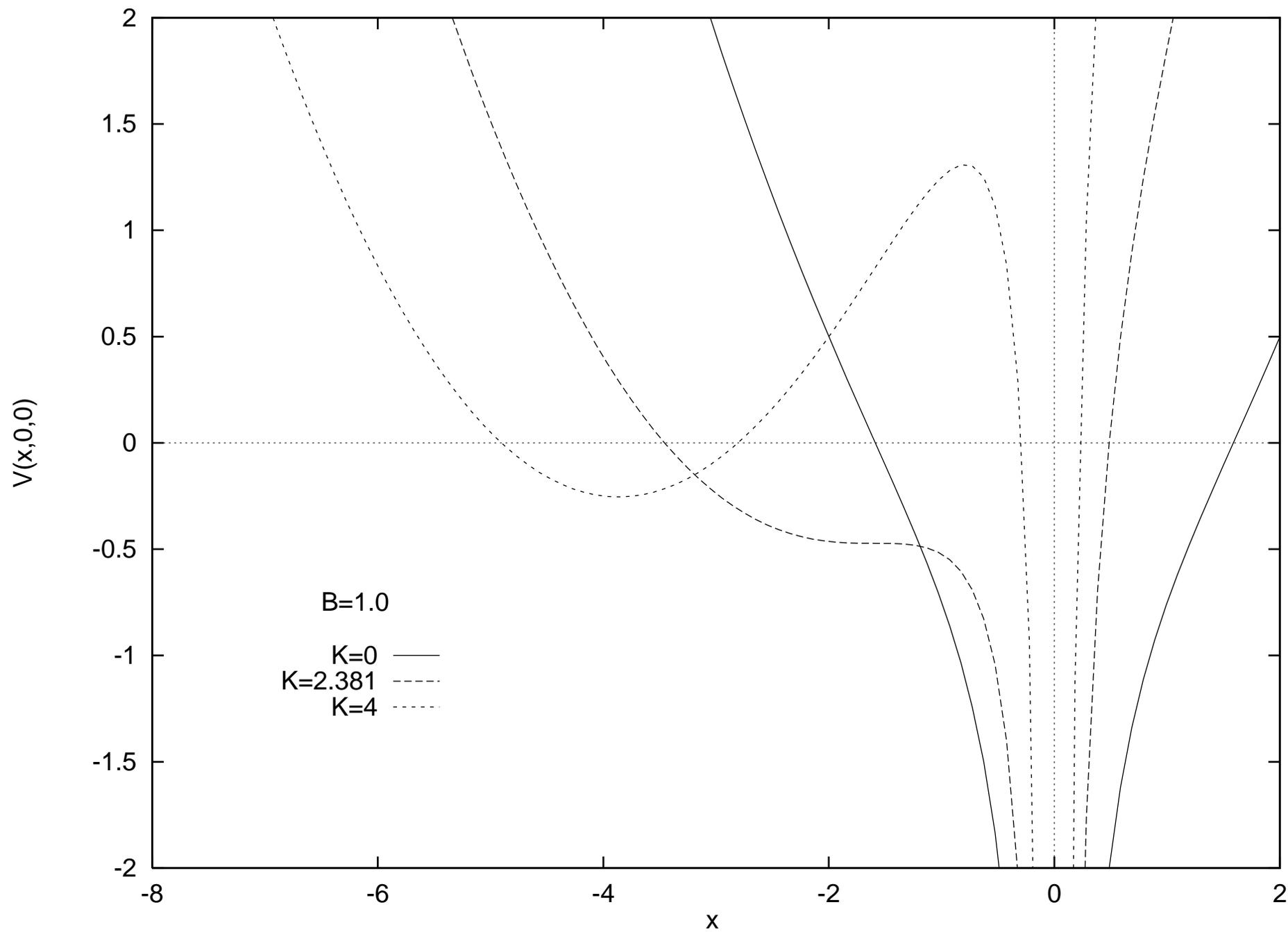
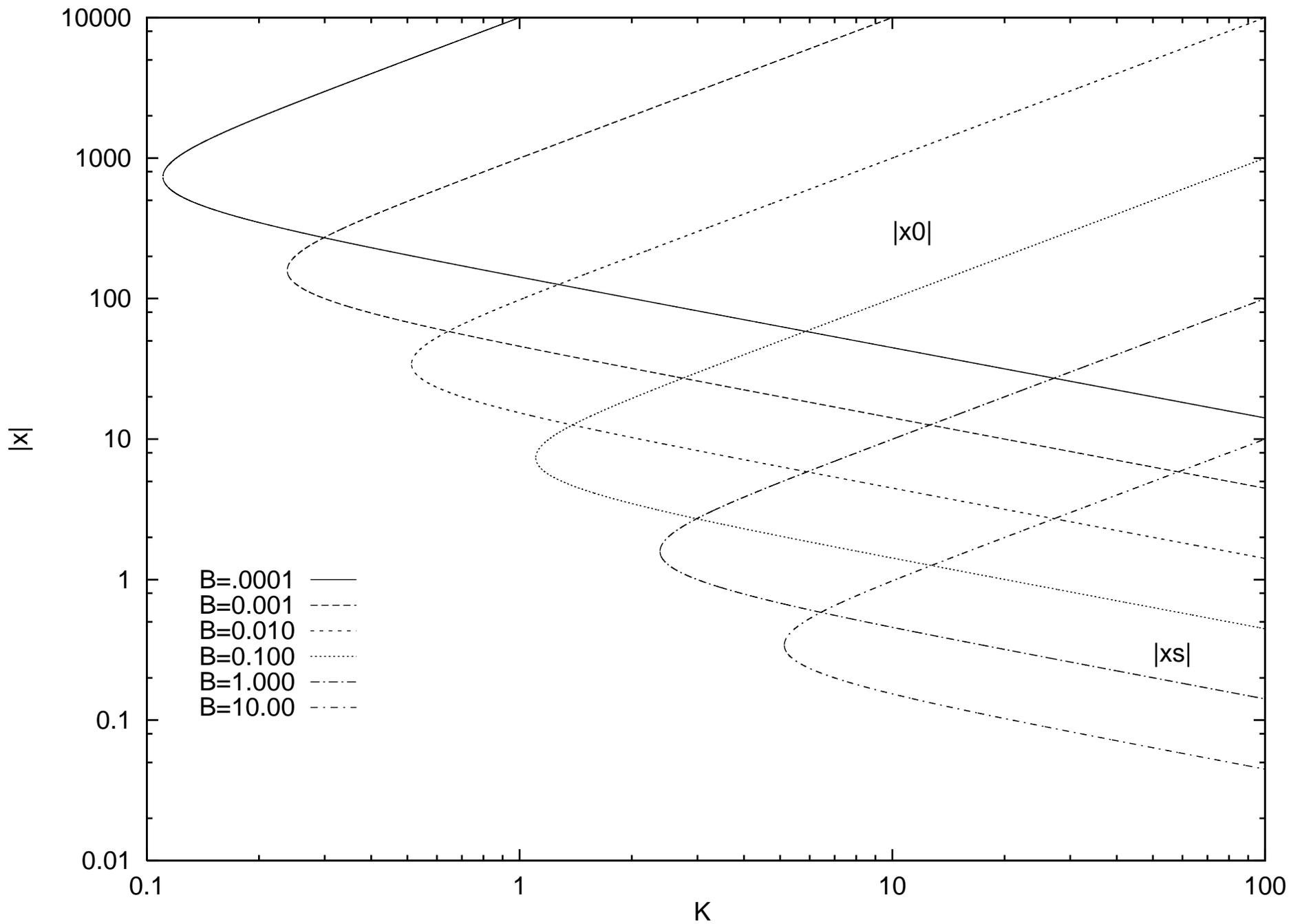
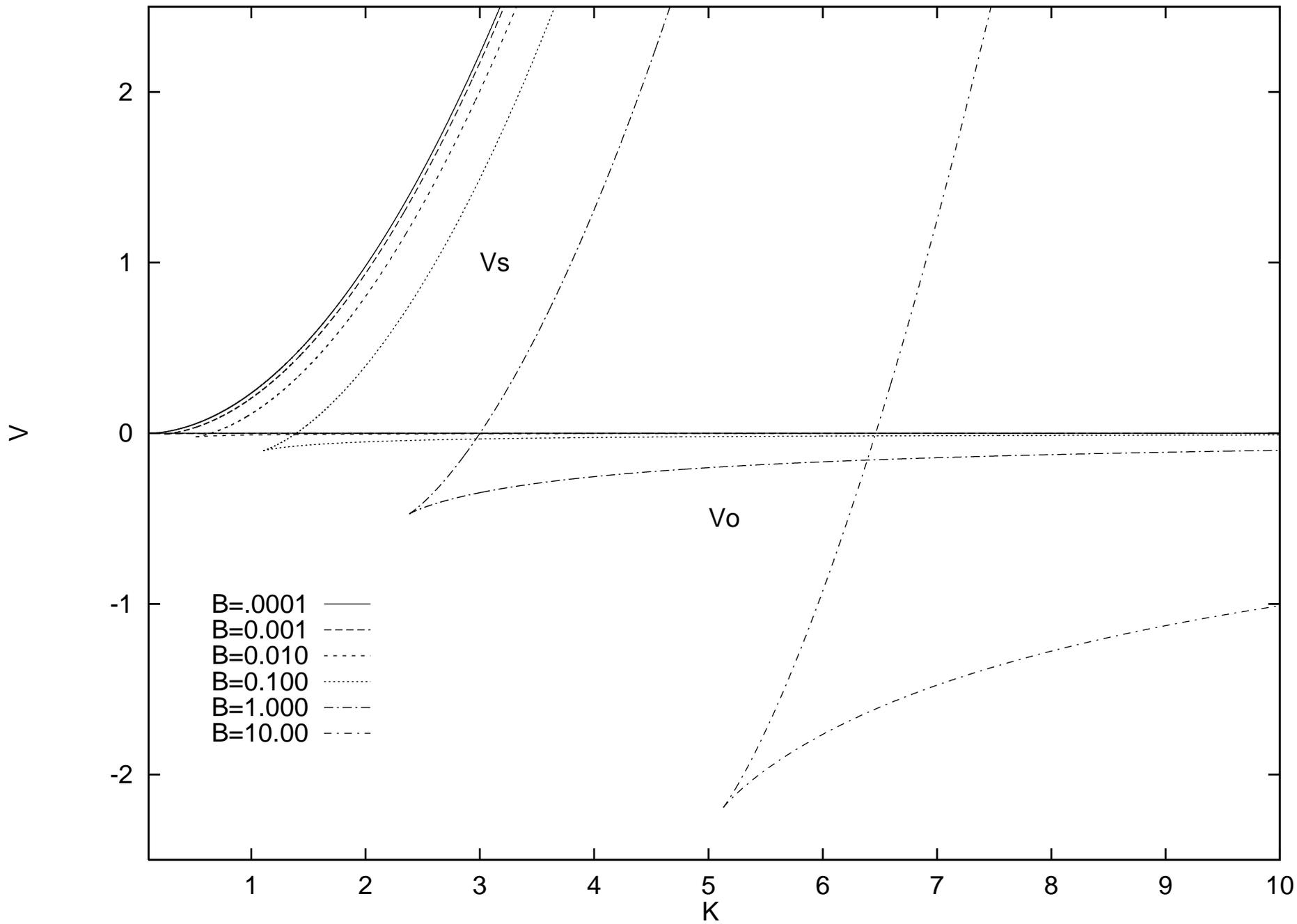
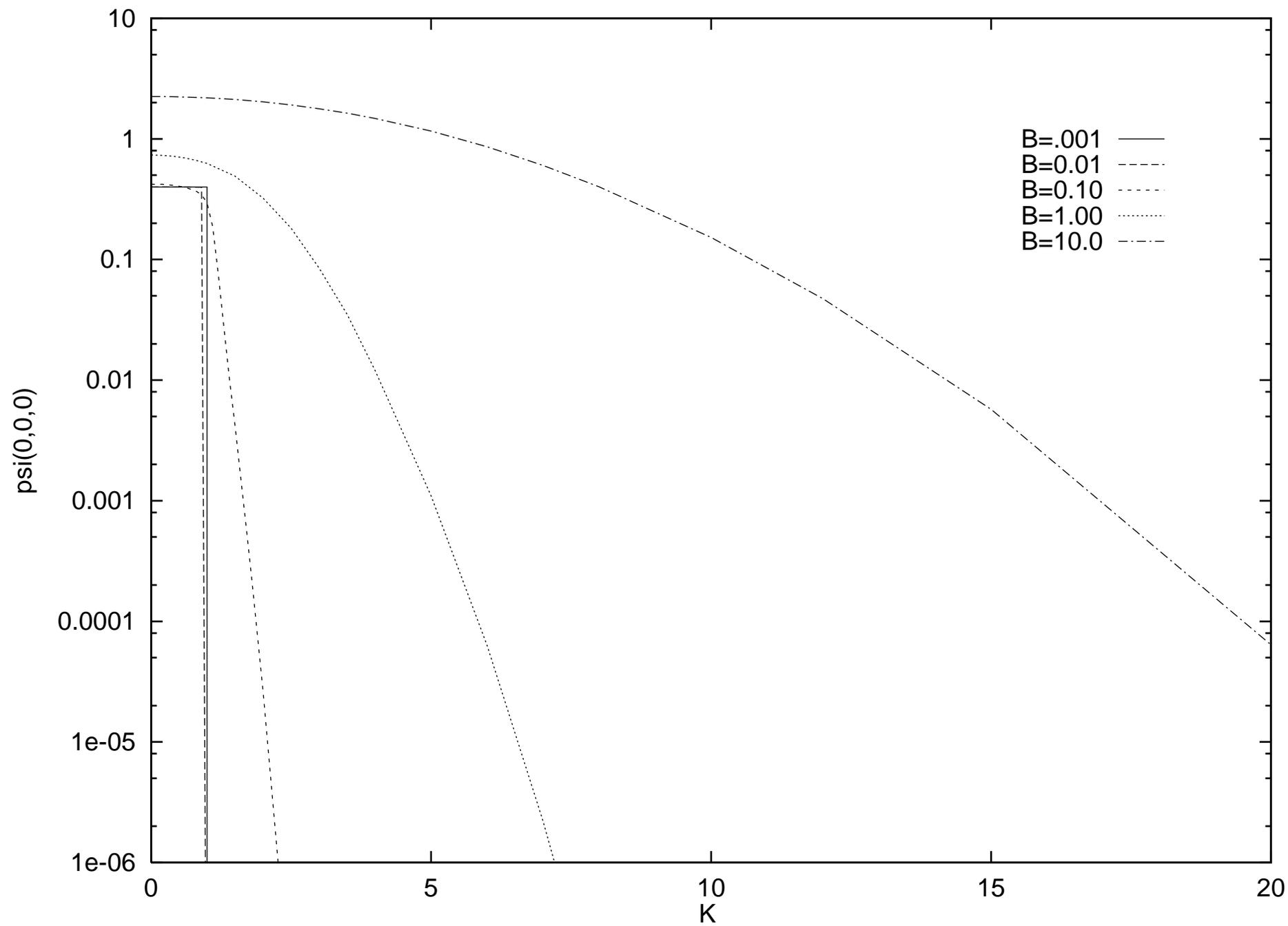
\end{document}